\documentstyle[aps,epsfig,multicol]{revtex}

\begin{document}
\draft
\title{Spectral and Transport Properties of  
$D$-Wave Superconductors With Strong Impurities} 

\author{Alexander Altland} 
\address{Institut f\"ur Theoretische Physik,
  Universit\"at zu K\"oln, Z\"ulpicher Strasse 77, 50937 K\"oln,
  Germany}
\maketitle                                 
\begin{abstract}
  One of the remarkable features of disordered $d$-wave
  superconductors is strong sensitivity of long range properties to
  the {\it microscopic} realization of the disorder potential.
  Particularly rich phenomenology is observed for the --
  experimentally relevant -- case of dilute distributions of isolated
  impurity centers.  Building on earlier diagrammatic analyses, the
  present paper derives and analyses a low energy effective field
  theory of this system.  Specifically, the results of previous
  diagrammatic $T$-matrix approaches are extended into the
  perturbatively inaccessible low energy regimes, and the long range
  (thermal) transport behaviour of the system is discussed.  It turns
  out that in the extreme case of a half-filled tight binding band and
  infinitely strong impurities (impurities at the unitary limit), the
  system is in a delocalized phase.
\end{abstract}

\begin{multicols}{2}
\narrowtext

\section{Introduction}
\label{sec:introduction}

Disorder in the cuprates is commonly attributed to the presence of dilute
distributions of strong local scatterers  -- e.g. $Zn$-atoms
replacing $Cu$ --
immersed into the two-dimensional matrices of $CuO_2$
planes\cite{fn1}.  While for conventional superconductors
details of the microscopic realization of the disorder potential are
of little, if any concern, the situation with anomalous
superconductors is different. In fact, {\it microscopic} aspects of
the impurity scattering have proven responsible for many of the
notorious difficulties encountered in developing quantitative theories
of disordered quasi-particles in $d$-wave superconductors:

{\it First}, the spectral characteristics of $d$-wave superconductors
polluted by dilute strong scatterers -- commonly dubbed 'unitary
scatterers' -- differ strikingly from those of systems with
continuously distributed disorder: While in the latter case, the
spectral density of quasi-particles categorically vanishes upon
approaching zero energy\cite{nersesyan94,senthil98}, it may {\it
  diverge} for unitary disorder\cite{pepin98}. This is demonstration
of the fact\cite{altland00.1,atkinson2} that the standard paradigm of
``insensitivity of long range properties of fermion systems to details
of the disorder'' is essentially violated in anomalous superconductors.

{\it Second}, conventional diagrammatic perturbation theory rules out
as a tool to quantitatively address the low-energy regime of the
system.  Based on self-consistent $T$-matrix (SCTA) approximation
schemes, early perturbative approaches to the
problem\cite{hirschfeld88,lee93} did not take the existence of
infrared-singular diagram
classes\cite{altland96,yashenkin01} into account.
These diagrams, similar in appearance to the diffusons and cooperons
of normal metals, reflect the existence of a Goldstone mode
linked to the particle-hole exchange symmetry of the system.  For not
too small energies, they can be brought under control by
summing up weak localization type corrections to the Gorkov Green
functions\cite{altland96,yashenkin01}.  However, for low energies these
corrections proliferate and perturbation theory produces unphysically
divergent results. For the same reason, diagrammatic theory cannot be
applied to assess the {\it transport} behaviour of the quasi-particle
system.

In contrast, field theoretical approaches to the
problem\cite{nersesyan94,senthil98,altland00.1} provide
the means for controlled inclusion of all low-energy soft modes.
However, {\it third}, the specific case of dilute impurity
distributions, commonly referred to as {\it Poisson distributed}
disorder, has been out of reach of previous fieldtheoretical
formulations. (This is because the standard construction route of
field theories of disordered fermions is based on models of
continuously, mostly Gaussian distributed disorder.)

To surmount these difficulties, a number of alternative approaches to
the problem have been developed. Building on an explicit
representation of the realization-specific Gorkov Green function,
Pepin and Lee (PL) \cite{pepin98} found the low energy single particle
density of states (DoS) $\nu(\epsilon)$ to be power-law divergent,
$\nu(\epsilon) \sim 1/|\epsilon \ln^2(|\epsilon|)|$. While this result
was obtained for the specific case of a half filled band, PL argue
that the band center singularity should be insensitive to not too
strong variations of the chemical potential. (Considering applications
to the cuprates, the stability issue is of prime relevance: At half
filling, or zero doping, the Cuprates are in an anti-ferromagnetic
phase, i.e. the very formation of a superconducting phase requires a
finite amount of dopands. This implies that strong sensitivity of the
band center singularity to deviations off half filling would impose severe
limitations on the practical relevance of the result.)

However, this latter statement is at variance with
numerical~\cite{atkinson2} and analytical~\cite{yashenkin01} findings
indicating that the band center singularity represents a highly
fragile phenomenon. As shown by Yashenkin et al.
(YAGHK)\cite{yashenkin01} the appearance of a zero energy divergence
is tightly linked to the formation of a second class of soft modes,
generated by the nesting symmetry of the half filled band and existing
in parallel to the standard particle-hole modes. Within perturbation
theory, these modes lead to the formation of {\it positive}
logarithmic (but not power law, like in PL) corrections to the low
energy DoS. Further, even slight deviations from the combined limit
(unitarity/half filling) generate a mass gap for the anomalous soft
modes and induce a crossover to the {\it negative} logarithmic
corrections characteristic for the standard particle-hole modes.

While the perturbative analysis reveals the existence of an anomalous
phase in the half filled system, it cannot reliably predict the
low-energy profile of the DoS, the critical properties of the
anomalous phase, and its transport properties. To shed more light on
the low energy phase of the Poisson disordered system in general,
Chamon and Mudry\cite{mudry00:_unitar} suggested a duality
transformation relating the limit of unitary scattering to the
opposite extreme of Gaussian, or Born disorder. The structure of this
transformation was suggested on the basis of variational equations
controlling the high-energy regime of the system.  However, since the
notorious soft modes do not at all enter the analysis of Chamon and
Mudry, the extensibility of the transformation into the fluctuation
dominated low energy regime remains speculative.

Building on these previous analyses, it is the objective of the
present paper to describe the Poisson-disordered $d$-wave
superconductor by fieldtheoretical methods.  
The generalized formalism will then be applied
to explore spectral and localization properties of the system, at and
away from half filling.  We begin our analysis with a brief review of
the general field theory approach to the $d$-wave superconductor, a
qualitative discussion of the extension to pointlike disorder, and a
summary of the results the generalized model predicts for spectral and
transport properties of the system (section
\ref{sec:qual-disc-results}).  The next two sections contain technical
details of the construction, first for generic pointlike disorder
(section \ref{sec:constr-field-theory}), then for the
unitary/half-filled limit (section \ref{sec:half-filled-band}). We
conclude in section \ref{sec:summary-discussion}.

\section{Qualitative Discussion and Results}
\label{sec:qual-disc-results}

As with normal  systems, the properties of
 disordered superconductors are to a large extend determined by symmetries.
Specifically, a time-reversal invariant spin-singlet superconductor
obeys the two relations,
\begin{eqnarray}
  \label{eq:1}
 {\cal C}&:&\hat H = - \sigma_2 \hat H^T \sigma_2,\nonumber\\
 {\cal T}&:& \hat H = \hat H^T,
\end{eqnarray}
where $\hat H$ is the Gorkov Hamilton operator and $\sigma_i$ are
Pauli-matrices acting in particle-hole space. The first of these
relations expresses the general particle-hole conjugation symmetry of
the spin-singlet superconductor while the second is time reversal. In
some sense the symmetry class identified through (\ref{eq:1}), termed
class $C$I in Ref. \cite{altland96}, plays a role analogous to the
standard orthogonal symmetry class of disordered normal systems.

For a conventional superconductor, the presence of a large
quasi-particle excitation gap implies that the symmetries (\ref{eq:1})
only marginally affect low-energy, or long-range system properties.
However, in a gapless environment, as realized, e.g., in the $d$-wave
superconductor, the situation is different. Here, the symmetry ${\cal
  C}$ entails the formation of long-lived particle-hole excitations: A
particle and a hole with relative energy difference $\epsilon$
interfere to form a slowly fluctuating composite mode $\Pi(\epsilon)$.
As $\epsilon\to 0$, the coherence between the particle and the hole
amplitude becomes perfect, implying infrared singular, or massless
behaviour of $\Pi(\epsilon)$.

The manifestation of these modes in observable low-energy properties
can efficiently be explored by field integral methods. A
fermion-replica field theory describing disordered superconductors of
class $C$I was first derived by Oppermann\cite{oppermann90,KrOp} and
later rediscovered, then within the context of $d$-wave
superconductivity, by Senthil and Fisher\cite{senthil98}.
Supersymmetric implementations of the theory, needed to explore the
perturbatively inaccessible infrared region, have been introduced in
Refs.\cite{altland00.1}.

All these models were derived for Gaussian distributed
disorder.  By manipulations similar to those employed in the
construction of the standard $\sigma$-models for systems of
Wigner-Dyson symmetry, the microscopic model was mapped onto an action
functional
\begin{equation}
  \label{eq:2}
Z = \int {\cal D}T \exp(-S[T]),  
\end{equation}
where $T$ is a field taking values in the group ${\rm Sp}(2r)$ ($r$:
number of replicas) for the replica-implementations, and in the
supergroup ${\rm OSp}(2|2)$ for supersymmetry. (For a precise
definition of this group, see Eqs. (\ref{eq:14}) and (\ref{eq:28})
below.)

As with the standard $\sigma$-models, the structure of the action of
this field theory is fixed by a few macroscopic system parameters.
Specifically, for a $2d$ gapless superconductor with diffusive
quasi-particle dynamics, the (bare) action has the form
\begin{eqnarray}
  \label{eq:3}
&&S[T] =\\
&&\hspace{.2cm}= {\pi \nu_0 \over 8}\int d^2 r
\left[ -D {\,\rm str\,}(\partial T
  \partial T^{-1}) 
  - 2i\epsilon {\,\rm str\,}
  (T+T^{-1})\right],\nonumber   
\end{eqnarray}
where $D$ is the diffusion constant and $\nu_0$ the DoS computed in
the self-consistent Born approximation (SCBA).  Importantly, the symmetry of
the fields $T$ -- which in turn is fixed by the physical symmetries
(\ref{eq:1}) -- does not permit other contributions to the low energy
action. In other words, the action is completely fixed by the value of
the two constants $D$ and $\nu_0$. Notice that for the $d$-wave
superconductor, and on the level of the bare theory, 
\begin{equation}
    \label{eq:29}
    D \nu_0 = {1\over \pi^2} {t^2 + \Delta^2 \over t\Delta}\equiv 4g_s,
\end{equation}
where $\Delta$ and $t$ are the order parameter and hopping strength,
respectively and $g_s$ is the spin conductance. I.e. the bare, or
Drude spin conductance does not depend on disorder
concentration\cite{lee93}.

Starting from the action (\ref{eq:3}), spectral and transport properties of the
quasi-particles can conveniently be explored. However, before turning
to the phenomenology of the system let us briefly discuss what changes
with the theory (\ref{eq:3}) if the disorder is Poissonian
distributed.  Certainly, the fundamental symmetries (\ref{eq:1}) are
insensitive to the specifics of the disorder distribution. Further,
the quasi-particle dynamics on intermediate time scales, larger than
the scattering time but smaller than any scale related to
quasi-particle localization, will continue to be diffusive. This
leaves us with two principal possibilities: Either there are some
fundamental reasons that exclude the existence of a local effective
action of the Poisson disordered system. Or, a low-energy
action exists {\it and} is bound to have the same structure as the
action (\ref{eq:3}).

The result of a technical derivation detailed in section
\ref{sec:constr-field-theory} will be that the second scenario is
realized, i.e. that the low-energy phase of the Poisson disordered
system is again described by (\ref{eq:3}).  The only difference to
the Gaussian case is that the value of the constant $\nu_0$ is now set
by the SC{\it T}A rather than by the SC{\it B}A density of states.

That both Gaussian and the Poissonian disorder map onto the field
theory (\ref{eq:3}) implies a number of predictions on the long range
behaviour of the system:
\begin{itemize}
\item[$\triangleright$] First, the quasi-particle DoS obtains as
\begin{equation}
  \label{eq:4}
  \nu (\epsilon) =  {\nu_0\over 8} {\rm \, Re\,}\langle {\, \rm
    str\,}((T+T^{-1}) {\cal P})\rangle_T, 
\end{equation}
where $\langle \dots \rangle = \int {\cal D}T e^{-S[T]}$ stands for
functional averaging and ${\cal P}$ is a constant projector
matrix\cite{fn2}.  In the most basic approximation to the functional
integral, fluctuations of the fields $T$ are neglected and one obtains
$\nu(\epsilon) = \nu_0$ in agreement with the SCBA/SCTA diagrammatic
result. Going beyond this approximation,
\item[$\triangleright$] a one-loop integration over fluctuations of
  the fields $T$ leads to a negative correction term\cite{senthil98},
  \begin{equation}
    \label{eq:43}
  \delta \nu(\epsilon) \sim -{1\over g_s} \ln\left(\Lambda 
      \sqrt{g_s\over \epsilon \nu_0}\right),
   \end{equation}
   where $\Lambda\sim a^{-1}$ is an ultraviolett cutoff and $a$ the
   lattice spacing. Eq.  (\ref{eq:43}) is in agreement with the
   results obtained by extended diagrammatic series
   summations\cite{yashenkin01}, i.e. analyses accounting for the
   presence of singular ladder diagrams. Notice that the magnitude of
   the quantum correction depends on the (disorder independent!)
   scale $g_s(t/\Delta)$: The smaller the anisotropy $t/\Delta>1$, the
   more important become quantum fluctuations. As concerns the 'mass
   parameter' $\nu_0 \epsilon$ truncating the logarithm, there is an
   important difference between the Gaussian and the Poissonian case.
   While for Gaussian disorder, $\nu_0$ {\it increases} quadratically
   with the strength of the impurity potential, $v$, it {\it
     decreases} as $v^{-1}$ for strong pointlike scatterers. (For
   quantitative expressions, see section \ref{sec:constr-field-theory}
   below.), i.e.  the small parameters of the perturbative approaches
   to the two limits scale reciprocally with the impurity strength,
   which is in agreement with the 'duality' hypothesis of
   Ref.\cite{mudry00:_unitar}.
 \item[$\triangleright$] For systems of finite size $L$ and energies
   smaller than the Thouless energy $\epsilon = {D\over L^2}$, spatial
   fluctuations of the fields $T$ are effectively frozen out. As with
   normal systems, a non-perturbative integration over the spatially
   constant zero mode $T_0$ obtains results compatible with RMT
   behaviour. Specifically, the low-energy spectral density scales as
  \begin{equation}
    \label{eq:7}
    \nu(\epsilon) \sim  {|\epsilon|\over \Delta},
  \end{equation}
  where $\Delta = (\nu_0 L^2)^{-1}$ is the single particle level
  spacing. 
\item[$\triangleright$] Turning to transport, it has been
  shown\cite{senthil98} that under renormalization, the conductance
  scales from its disorder-indepedent bare value to zero, with a
  $\beta$-function $\beta = 1/(4\pi^3 g_s^2)$. This implies that for
  large systems quasi-particles should be localized on the scale of
  some localization length $\xi\sim \exp (4\pi^2 g_s)$; In the
  thermodynamic limit the system is a (spin) insulator. Similarly,
\item[$\triangleright$] in the limit $L\gg \xi$, mechanisms of quantum
  localization and spectral repulsion conspire to manufacture a
  spectral gap at zero energy: Consider a quasi-particle at energies
  $\epsilon < D/\xi^2$, corresponding to time scales larger than the
  time needed to diffusively explore a localization volume $\sim
  \xi^2$. Confined to stay within the localization volume, the
  quasi-particle will exhibit ergodic behaviour, i.e.  the dynamics
  will become random matrix like. One thus expects\cite{senthil98} that
   \begin{equation}
     \label{eq:8}
     \nu(\epsilon) = {|\epsilon|\over \Delta_\xi},\qquad |\epsilon| <
     {D\over \xi^2},
   \end{equation}
   where $\Delta_\xi = (\nu_0 \xi^2)^{-1}$ is the level spacing of a
   single localization volume. (While for the two-dimensional
   superconductor Eq.~(\ref{eq:8}) has the status of a conjecture,
   similar predictions for quasi one-dimensional quantum wires of
   non-Wigner-Dyson symmetry were confirmed by transfer
   matrix methods\cite{brouwer00_off,altland01}.)
\end{itemize}
Summarizing, we find that for both Gaussian and Poisson disorder the
system scales to a spin-insulator phase. While the DoS's obtained for
the two types of disorder are qualitatively similar, the extension of
regimes with different spectral properties scales reciprocally with
the disorder strength.

While these results apply to the case of generic disorder, there is
one particular point in the phase diagram where drastically different
behaviour is observed: In the combined limit half-filling/unitary
scattering, the zero energy DoS exhibits a power law {\it
  divergence}\cite{pepin98}.  In YAGHK the peculiar properties of this
limit, henceforth referred to as the 'unitary limit' for brevity,
were related to the formation of a second class of soft modes,
existing in parallel to the particle-hole modes discussed above. It
was shown that that perturbative inclusion of these modes leads to
{\it positive} logarithmic corrections to the DoS. However, as with
the generic particle-hole modes discussed above, no safe conclusions
on the profile of the DoS can be drawn on the basis of pure
perturbation theory.

Below we will show that the emergence of a new class of soft modes is
indicative of critical behaviour. The system at unitarity {\it and}
$\epsilon=0$ sits on a critical line (parameterized through the ratio
$t/\Delta$.) One has to emphasize that this fixed line lies outside
the phase diagram of 'real' high $T_c$ materials; cuprates at half
filling are anti-ferromagnets rather than superconductors.  However,
in the {\it vicinity} of the critical regime spectral and transport
observables will exhibit anomalous scaling behaviour traces of which
may (or may not) be observable in real systems.

But what, then, is the origin of the exotic behaviour? In
YAGHK the formation of the anomalous soft modes was traced back to the
nesting symmetry of the half filled band. The $2\times 2$ matrix
Gorkov Hamiltonian function of the {\it clean} system obeys the
symmetry
\begin{equation}
  \label{eq:30}
  {\cal N}: \hat H_0({\bf k}) = - \hat H_0 ({\bf k} + {\bf Q}),  
\end{equation}
where ${\bf Q}=(\pi,\pi)$ is the nesting vector and a square lattice
with unit lattice spacing is understood. Notice that what matters in
the description of spectral/transport properties is not the
Hamiltonian but the Green function $\hat G = (\epsilon + \mu \sigma_3
- \hat H)^{-1}$ ($\sigma_i$, $i=1,2,3$: Pauli matrices in
particle-hole space), i.e. for finite energy arguments, $\epsilon$, or
chemical potentials, $\mu$, the symmetry ${\cal N}$ is effectively
broken.

Now, we are dealing with a system where translational invariance is
broken by disorder, i.e.  it seems more natural to express the
symmetry ${\cal N}$ in real space. Noticing that translation by ${\bf
  Q}$ amounts to multiplication of every other site of the bipartite
lattice by $-1$, we get 
\begin{equation}
  \label{eq:31}
  {\cal N}: [\hat H_0,\sigma_3^{\rm sl}]_+=0,
\end{equation}
where $[\;,\;]_+$ is the anti-commutator and the action of
$\sigma_3^{\rm sl}$ is to multiply all sites belonging to one of the
two sublattices $A$ and $B$ nested into the bipartite host lattice
$A\cup B$ by $-1$. 

The representation (\ref{eq:31}) will not only turn out to be
technically convenient, it also provides the clue as to the special
role played by unitary scatterers. Evidently, a generic disorder
potential superimposed on the clean Hamiltonian $\hat H_0$ will not
respect the symmetry ${\cal N}$. With unitary, i.e.  'infinitely
strong' scatterers, however, the situation is different. In a sense,
such impurities burn local holes into the system. More technically,
the potential $v$ created by unitary scatterers is strong enough to
locally force the amplitude of wave functions in the physical Hilbert
space of the system to zero.  Now, multiplication of zero by $\pm 1$
does not change anything which is why the symmetry ${\cal N}$ is
effectively {\it left intact} by unitary scatterers.

In which way does the symmetry ${\cal N}$ affect the long range
behaviour of the system? Temporarily forgetting about
superconductivity, let us begin by briefly recapitulating the role of
${\cal N}$-symmetries in {\it conventional} metallic systems. Building
on early work by Wegner and Oppermann\cite{oppermann79},
Gade\cite{gade93} realized that sublattice symmetries drastically
affect the properties of  disordered systems
{\it in the middle} of the tight binding band.  Specifically, (i) the
middle of the band supports a delocalized phase and (ii) the DoS
diverges as a power law $\nu(\epsilon) \sim |\epsilon|^{-1} e^{-\kappa
  \sqrt{-\ln(|\epsilon|)}}$, with some non-universal parameter
$\kappa$, upon approaching the band center.  These phenomena reflect
the fact that the $\epsilon=0$ system with ${\cal N}$-symmetry belongs
to a non-Wigner-Dyson symmetry class, viz.  class $A$III in the
classification of Ref.\cite{zirnbauer96}. Breaking this symmetry,
finite values of the parameters $\epsilon$ or $\mu$ induce a crossover
to the Wigner-Dyson unitary class (or class $A$), i.e. the symmetry
class of generic electronic systems with broken time reversal
invariance.

To obtain these results Gade described the normal metal sublattice
system in terms of a (boson replica) effective field theory, which was
then subjected to a renormalization group analysis. For future
reference we note that this theory has a supersymmetric extension
defined through $Z=\int {\cal D}T \exp(-S[T])$ and
\begin{eqnarray}
  \label{eq:32}
  &&S[T] = \nonumber \\
&&\hspace{.2cm}={\pi \nu_0 \over 8}\int d^2 r 
 \left[ -D{\,\rm str\,}(\partial T
  \partial T^{-1}) 
  - 2 i \epsilon {\,\rm str\,}
  (T+T^{-1})\right]+\nonumber\\
&&\hspace{.5cm} + c \int d^2r \left[{\rm str}(T\partial T^{-1})\right]^2,
\end{eqnarray}
where $c$ is a non-universal disorder-dependent constant and the
fields $T$ take values in the supergroup ${\rm GL}(1|1)$ (i.e. the
group of invertible two-dimensional supermatrices.)

While Gade obtained the field theory (\ref{eq:32}) (or rather its
boson replica analog) for the specific model of electron hopping on a
bipartite lattice, it is now understood that there is a more general way
of looking at the problem (cf., e.g., the discussion in Ref.
\cite{altland99:NPB_flux}). Field theories like (\ref{eq:32})
generally describe the long range behaviour of Hamiltonians of
symmetry class $A$III. A Hamiltonian belongs to this symmetry class if
it can be represented in a {\it block form},
\begin{equation}
  \label{eq:44}
  \hat H = \left(\matrix{& Z \cr Z^\dagger&}\right),
\end{equation}
where $Z$ are arbitrary matrices.  (Indeed, the symmetry
(\ref{eq:31}) states that the Hamiltonian assumes a block
off-diagonal form when represented in a sublattice basis.) 

The generalized interpretation helps to understand what happens with
the sublattice system when the superconductive aspects of the problem
are switched on. Indeed, it is straightforward to show  (cf.  Appendix
\ref{sec:block-repr-superc}) that (i) the superconductor
sublattice Hamiltonian can be transformed to a representation
(\ref{eq:44}) and (ii) that there are no other symmetries in the
problem.  This implies that the superconductor with ${\cal
  N}$-symmetry is again described by the low-energy effective action
(\ref{eq:32}), a claim to be verified by explicit construction in
section \ref{sec:half-filled-band}. There is one {\it difference} to
the previously discussed case, viz. that the fields $T$ now take values
in the larger group ${\rm GL}(2|2)$ (i.e.  extra degrees of freedom
are needed to accommodate the superconductor symmetries (\ref{eq:1}).)
However, as far as the long range properties of the system are
concerned, an extension of the field space to ${\rm GL}(n|n)$ (here
$n=2$) is inconsequential: As shown by renormalization group methods
in Ref.\cite{gade93},
\begin{itemize}
\item[$\triangleright$] for zero excitation energy, $\epsilon=0$, the
  system is critical.
\item[$\triangleright$] Upon approaching zero energy, the DoS {\it
    diverges} as 
  \begin{equation}
    \label{eq:39}
    \nu(\epsilon) \sim {e^{-\kappa \sqrt{-\ln(|\epsilon|)}}\over
      |\epsilon|}, 
  \end{equation}
  where $\kappa$ is some non-universal, disorder dependent scale that
  cannot be reliably determined from the RG analysis. For
  sufficiently small energies, $\epsilon< \exp (-\kappa^2)$, the
  DoS scales in a near power-law manner, $\nu(\epsilon) \sim
  |\epsilon|^{-1}$. 
\item[$\triangleright$] The conductance does not renormalize, i.e. the
  system is a (thermal) metal.
\item[$\triangleright$] All three, finite values of $\epsilon$, the
  chemical potential $\mu$, and deviations off unitarity, $v^{-1}>0$,
  represent relevant perturbations that drive the system towards the
  generic superconductor symmetry class $C$I.
\end{itemize}

While the first three of the characteristics listed above can directly
be inferred from Gade's analysis, the last needs some 
clarification. As shown in section \ref{sec:half-filled-band},
 deviations off the unitary limit lead to the
appearance of an extra operator
\begin{equation}
  \label{eq:33}
  S_{\rm sb} = \Gamma \int d^2 r {\, \rm str\,}(T T^s + (TT^s)^{-1}),
\end{equation}
to be added to (\ref{eq:32}). Here, $\Gamma=\Gamma(\mu, v^{-1})$ is a
symmetry breaking parameter whose explicit dependence on $\mu$ and
$v^{-1}$ is given in Eqs. (\ref{eq:40}) and (\ref{eq:42}) below.
Further, $X^s$ denotes a kind of generalized matrix transposition.
More important than the explicit definition of this operation (see Eq.
(\ref{eq:14}) below) is that finite values of $\Gamma$ enforce $T^{s}=
T^{-1}$ which is the defining relation of the orthosymplectic group
${\rm OSp}(2|2) \subset {\rm GL}(2|2)$. In other words, the action
$S_{\rm sb}$ induces a crossover between the extended symmetry at the
unitary limit ($A$III) and the generic symmetry class
$C$I\cite{fn3}.  When embedded into an RG analysis, the operator
$S_{\rm sb}$ represents a relevant perturbation with scaling dimension
2 (by elementary power counting.) I.e.  deviations off the unitary limit
rapidly increase under the RG operation and drive the system towards
the phase of the generic superconductor.

\begin{figure}[hbt]
\centerline{\epsfxsize=3.5in\epsfbox{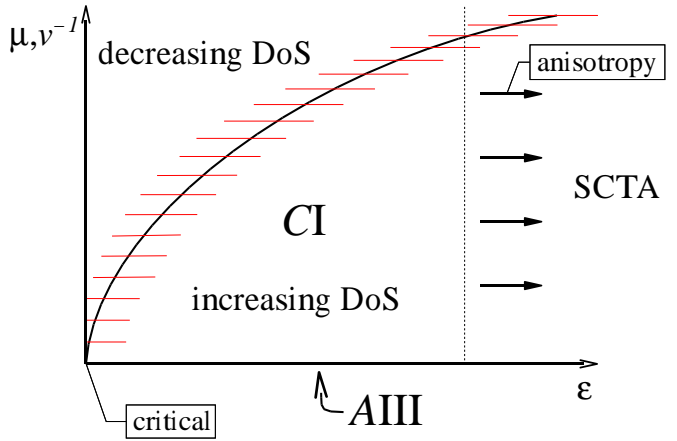}}\vspace{0.5cm}
\caption{Qualitative diagram of regimes with different behaviour of
  the DoS $\nu(\epsilon)$. Horizontal axis: energy, $\epsilon$;
  vertical axis: ${\cal N}$-breaking by finite chemical potentials
  $\mu$ and/or non-infinite scattering strengths $v^{-1}$. On the abscissa the
  system is in class $A$III, everywhere else in $C$I. For large values
  of $\epsilon$ the DoS is well described by SCTA. The energy below
  which quantum corrections begin to play a role, increases with the
  anisotropy parameter $\Delta/t$.  Depending on whether the product $
  \epsilon \nu_0$ is smaller or larger than the symmetry breaking
  parameter $\Gamma(\mu , v^{-1})$ (the crossover region indicated by
  the solid line), quantum corrections to the DoS are negative or
  positive. At the origin of the diagram $(\Gamma,\epsilon)=(0,0)$,
  the system is critical and the DoS diverges.}
\label{fig:CI_AIII}
\end{figure}\vspace{0.5cm}

We finally discuss how these findings compare to the results of
previous analyses, notably PL and YAGHK. To leading order, the
divergence of the DoS, (\ref{eq:39}), agrees with the result of PL.
There are, however, some discrepancies on the level of logarithmic
corrections to powerlaw scaling (PL find $\nu(\epsilon) \sim
1/|\epsilon| (\ln(|\epsilon|)^2$.)  whose origin we do not understand.
The positive logarithmic corrections to the SCTA-DoS found in YAGHK
are perturbative precursors of the full powerlaw divergence. As a new
result, we find that transport in the band center is metallic.  It is
tempting to interpret this metallic behaviour as a consequence of
resonant tunneling transport in the narrow impurity band created by
unitary scattering centers.  However, this picture is purely
speculative. Finally, we disagree with the assertion of PL, that the
divergence survives generalization to finite values of $\mu$. Using RG
language, we rather find that $\mu$ acts like a (dynamically
increasing) mass scale terminating the RG flow for small enough
$\epsilon$. More specifically, one expects (cf.  Fig.
\ref{fig:CI_AIII}) that for large values of $\epsilon$ (i) the DoS
will be given by the SCTA result. Perturbative quantum corrections to
the large energy asymptotics (ii) scale logarithmically with
$\epsilon$ and inversely proportional to the anisotropy parameter
$t/\Delta \gg 1$. In an intermediate energy regime, RG summation over
all quantum interference contributions leads to the increasing DoS
profile (\ref{eq:39}).  Finally, (iii) below a crossover scale set by
$\nu_0 \epsilon/\Gamma(\mu,v^{-1})=1$, the DoS follows the {\it
  decreasing} profile of the superconductor of class $C$I.
Non-monotonous behaviour of this type has indeed been observed
numerically in YAGHK. Whether or not traces of the intermediate regime
(ii), or even of the scaling law (\ref{eq:39}), might be observable
experimentally cannot be decided by the present approach. (Mentioning
experiment, one should also notice that quasi-particle
{\it interactions} act as an additional ingredient spoiling sublattice
symmetries. I.e. the Hartree-Fock potential seen by the quasi-particles
 is local in real space  and, therefore, violates (\ref{eq:31}).)

 This concludes our qualitative discussion. Starting from a
 microscopic lattice Hamiltonian, the subject of the next two sections
 will be the construction the field theories (\ref{eq:3}) and
 (\ref{eq:32}). Basically, this will amount to an embedding of
 SCTA-summation techniques into a field integral approach. (Readers
 not interested in technicalities may skip these sections.)

\section{Construction of the Field Theory}
\label{sec:constr-field-theory}

Consider the quasi-particle Hamiltonian of a two-dimensional
lattice $d$-wave superconductor,
\begin{eqnarray}
  \label{eq:9}
  \lefteqn{\hat H = \sum_{ij}\Psi^\dagger_i\left( H^{0}_{ij} +
      H^{{\rm dis}}_{ij}\right)\Psi_j,}\\ 
&&\hspace{.5cm}\hat H^0_{ij} =  \xi_{ij} \sigma_3 +
\Delta_{ij} \sigma_1,\nonumber\\
&&\hspace{.5cm}\hat H^{{\rm dis}}_{ij} = V^{\rm p}_i \delta_{ij} \sigma_3+
V^{{\rm g}2}_{ij} \sigma_3 +
V^{{\rm g} 1}_{ij} \sigma_1\nonumber,
\end{eqnarray}
where $\xi_{ij}=t_{ij}-\mu$,
\begin{equation}
  \label{eq:10}
  \Psi_i = \left(\matrix{c_{i\uparrow}\cr c_{i\downarrow}^\dagger}\right),
\end{equation}
is a spinor comprising a spin-up hole and a spin-down particle, and
$\sigma_i$ are Pauli matrices acting in particle-hole space. The clean
part of the Hamiltonian, $\hat{H}^0$, contains the chemical potential,
$\mu$, a nearest neighbour hopping matrix element, $t_{ij}$, and the
$d$-wave order parameter, $\Delta_{ij}$. The Fourier components of
these operators are given by
\begin{eqnarray}
\label{eq:15}
\xi(k) &\equiv & t (\cos k_x + \cos k_y)-\mu ,\nonumber \\
\Delta(k) &=& \Delta (\cos
k_x - \cos k_y),  
\end{eqnarray}
where $t$ and $\Delta$ define the strength of hopping and  order
parameter, respectively. 

The stochastic part of the Hamiltonian, $\hat H^{\rm dis}$, accounts
for two different types of disorder. First, the potential generated by
$N$ randomly placed impurities is represented through
\begin{equation}
  \label{eq:36}
  V_i^{\rm p}=\sum_{j=1}^N v({\bf r}_i - {\bf R}_j), 
\end{equation}
where $\{{\bf R}_j\}$ are the coordinates of the impurity centers.
Configurational averaging over $V_i^{\rm p}$,
$$
\langle  \dots \rangle_{\rm p} \equiv \prod_{j=1}^N {1\over L^2}\int
d^2 R_j \;(\dots),
$$
will amount an independent, or Poissonian average over impurity
coordinates. 

Second, the random operators  $V^{{\rm g}a}$, $a=1,3$
model the presence of  residual statistical fluctuations
superimposed on the hopping and the order parameter amplitudes,
respectively. For convenience, we assume these amplitudes to
be Gaussian distributed with zero mean and variance,
\begin{equation}
  \label{eq:11}
  \langle V_{ij}^{{\rm g}a}  V_{i'j'}^{{\rm g}a'}\rangle_{\rm
    g} = g \delta_{ii'}\delta_{jj'} \delta^{aa'}.
\end{equation}
The motivation behind introducing Gaussian bond disorder in addition
to the Poisson impurity system is not only to make the modeling more
'realistic'. Below we will employ the Gaussian disorder average as a
vehicle to introduce the soft fields $T$. Averaging over the impurity
system will then be done in an after-step\cite{fn4}. (Notice that for
the purposes of the construction, the variance $g$ can be assumed to
be infinitesimally weak.)

To prepare the construction of a field theory of this system, we change
to an 'off-diagonal' representation, where the role of the Pauli
matrix $\sigma_3$ is taken over by $\sigma_1$. This is done by a
unitary transformation $\hat H \to \exp(-i{\pi\over 4} \sigma_1) \hat
H \exp(i{\pi\over 4} \sigma_1)$, after which
\begin{eqnarray*}
  &&\hspace{.5cm}\hat H^0 =  \hat \xi \sigma_2 +
\Delta \sigma_1,\nonumber\\
&&\hspace{.5cm}\hat H^{{\rm dis}} = (V^{\rm
  p}_i+V^{{\rm g}_2}) \sigma_2+
V^{{\rm g}1} \sigma_1\nonumber.
\end{eqnarray*}
The new representation changes the appearance of the basic symmetries
(\ref{eq:1}), viz.
\begin{eqnarray}
  \label{eq:1a}
 {\cal C}&:&\hat H = - \sigma_2 \hat H^T \sigma_2,\nonumber\\
 {\cal T}&:& \hat H = \sigma_1 \hat H^T \sigma_1.
\end{eqnarray}
We next introduce a supersymmetric partition function
\begin{eqnarray}
\label{eq:12}
&& Z[\hat J] = \int {\cal D}(\bar \psi,\psi) \exp\left(i \bar \psi
    \left[ \epsilon - \hat H + \hat J\right]\psi \right),
\end{eqnarray}
from which disorder averaged matrix elements of the retarded Gorkov
Green function can be obtained by differentiation with respect to
elements of the source matrix $\hat J$.  In (\ref{eq:12}), ${\,\rm Im
  \,}\epsilon >0$ and
$$
\psi=\left(\matrix{\psi^{\rm b}\cr \psi^{\rm f}}\right)
$$
is a four-component superfield whose complex commuting
(anticommuting) components $\psi^{\rm b}$ ($\psi^{\rm f}$) carry a
particle-hole structure like in (\ref{eq:10}). While convergence of
the integral requires $\bar \psi^{\rm b} = \psi^{\rm b \dagger}$, the
Grassmann fields $\bar\psi^{\rm f}$ and $\psi^{\rm f}$ are
independent.  For transparency, the lattice index summation is
suppressed in the notation.  Similarly,  the source field $\hat J$
will be set to zero in much of our further
discussion.

To fully account for {\it both} symmetries (\ref{eq:1}), it is
necessary to subject the functional integral to a doubling, or
'charge conjugation' operation. Using that
\begin{eqnarray*}
\lefteqn{  \bar \psi \hat H \psi = {1\over 2}\left( \bar \psi \hat H
    \psi +
 \psi^T \sigma_3^{\rm bf} \hat H^T \bar
 \psi^T\right)\stackrel{\protect (\ref{eq:1})}{=}}\\ 
&&\stackrel{\protect (\ref{eq:1})}{=} {1\over 2}\left( \bar \psi \hat H
    \psi -
 \psi^T \sigma_2\otimes \sigma_3^{\rm bf} \hat H \sigma_2 \bar  \psi^T\right),
\end{eqnarray*}
the functional integral can be re-written as,
\begin{eqnarray*}
&& Z[0] = \int {\cal D}\Psi \exp\left({i\over 2} \bar \Psi
    \left[ \epsilon \sigma_3^{\rm cc} - \hat H\right]\Psi \right),
\end{eqnarray*}
where the eight-component fields
\begin{equation}
  \label{eq:5}
  \Psi = \left(\matrix{\psi \cr \sigma_2 \bar \psi^T}\right),\qquad
  \bar \Psi = \left(\bar \psi, - \psi^T \sigma_2 \otimes \sigma_3^{\rm
      bf}\right).
\end{equation}
Here, the matrices $\sigma_i$, $\sigma_i^{\rm bf}$, and
$\sigma_i^{\rm cc}$ are Pauli matrices acting in particle-hole,
boson-fermion, and in the newly introduced charge-conjugation space,
respectively. 
The fields $\Psi$ and $\bar \Psi$ are related to each
other by
\begin{equation}
  \label{eq:6}
  \bar \Psi = \Psi^T \sigma_2 \otimes \tau, \qquad \Psi = - \sigma_2
  \otimes \tau \bar \Psi^T,
\end{equation}
where 
\begin{equation}
  \label{eq:13}
  \tau = E_{11}^{\rm bf} \otimes (i \sigma_2^{\rm cc}) + E_{22}^{\rm
    bf} \otimes \sigma_1^{\rm cc}
\end{equation}
and $E_{ij}^{\rm bf}$ is a matrix acting in ${\rm bf}$-space and
containing zeros everywhere save for a unity at position $ij$.
Fortunately, all we need to know about the  $\Psi$'s is
encapsulated by (\ref{eq:6}), i.e. there will be no need
to explicitly unravel the entangled eight component structure of these
fields. 

Continuing with the construction, we next average the functional
integral over {\it Gaussian} disorder and decouple the resulting
$\Psi^4$-interaction by means of a Hubbard-Stratonovich
transformation:
\begin{eqnarray}
\label{eq:20}
\lefteqn{  Z_{{\rm g}}[0]\equiv \langle Z[0]\rangle_{\rm g}= \int {\cal
    D}(Q_0,Q_3)
e^{-{1\over g} \sum_{ij} {\rm \,str\,}(Q_{0,ij}^2+Q_{3,ij}^2)}
\times}\nonumber\\
&&\hspace{0.2cm}\times \int {\cal D}\Psi\exp\left({i\over 2} \bar \Psi
    \left[ \epsilon \sigma_3^{\rm cc} + 4 \hat Q_0 + 4 i \hat Q_3
      \sigma_3 - \hat       H^{\rm p}\right]\Psi \right)=\nonumber\\
&&\hspace{1cm}=\int {\cal
    D}(Q_0,Q_3) e^{-{1\over g} \sum_{ij}
{\rm      \,str\,}(Q_{0,ij}^2+Q_{3,ij}^2)} \times\nonumber\\
&&\hspace{0.5cm}\times \exp -{1\over 2}{\,\rm str\; ln\,}\left(
 \epsilon \sigma_3^{\rm cc} + 4 \hat Q_0 + 4 i \hat Q_3 \sigma_3 - \hat
      H^{\rm p}\right),
\end{eqnarray}
where $\hat H^{\rm p}= \hat H\big|_{V^{\rm g}=0}$ contains only
Poisson disorder and $Q_{a,ij}$, $a=0,3$ are four component
supermatrices living on the non-directed links $(ij)$ of the lattice
and acting in ${\rm bf}$ and ${\rm cc}$ space. Finally,
$$
(\hat Q_a)_{ij} \equiv \delta_{ij} {1\over 4}\sum_{l\in N_i} Q_{a,il},
$$
where $N_i$ is the set of four nearest neighbours of site $i$.

We next subject the functional integral to a stationary phase
approximation. Temporarily neglecting both the energy increment
$\epsilon$ and the impurity potential one finds that the variational
equations ${\delta S\over \delta Q_a} = 0 $ are solved by the
matrix-diagonal and spatially uniform configuration $(\bar Q_0,\bar
Q_3)=i{\kappa\over 4}( \sigma_3^{\rm cc},0)$.  Here, $\kappa(g)$ is some
function of the disorder strength which vanishes monotonously in the
limit $g \searrow 0$.

As usual with non-linear $\sigma$-models, the diagonal configuration
is but one representative of an entire manifold of solutions of the
mean field equation. To identify this manifold, we imagine our
diagonal configuration substituted into the second line of
(\ref{eq:20}) and explore what happens when the $\Psi$-field is
subjected to a transformation (matrix structure in ${\rm ph}$-space),
\begin{eqnarray*}
\Psi \to \left(\matrix{T_1&\cr&T_2}\right)\Psi,\qquad
\bar \Psi \to \bar \Psi \left(\matrix{\bar T_2&\cr&\bar T_1}\right).
\end{eqnarray*}
Here, $T_1,\dots,\bar
T_2$ are spatially constant supermatrices of dimension
four subject to the condition (to ensure compatibility with (\ref{eq:6}))
\begin{equation}
  \label{eq:14}
  \tau^{-1} T_i^T \tau \equiv T_i^s = \bar T_i.
\end{equation}
Remembering that $\hat H$ is off-diagonal in ${\rm ph}$-space, we find
that the transformation leaves the Hamiltonian invariant provided that
\begin{equation}
  \label{eq:28}
  \bar T_i = T_i^s = T_i^{-1},  
\end{equation}
which is the defining relation of the supergroup ${\rm OSp}(2|2)$.

We thus find that, notwithstanding the presence of the Poisson
impurity potential in $H^{\rm p}$, the action has the continuous group
${\rm OSp}(2|2)\otimes { \rm OSp}(2|2)$ as an invariance manifold.
While these transformations commute with the Hamiltonian, they change
the diagonal saddle point configurations according to
$$
i \kappa \left(\matrix{\sigma_3^{\rm cc}&\cr & \sigma_3^{\rm cc}}\right) \to 
i \kappa \left(\matrix{T_2^{-1}\sigma_3^{\rm cc}T_1&\cr &
    T_1^{-1}\sigma_3^{\rm cc}T_2}\right). 
$$
This equation identifies the set of all field configurations
$T\equiv T_2^{-1} \sigma_3^{\rm cc} T_1 \in {\rm OSp}(2|2)$ as the
Goldstonemode manifold of the model. For fixed $T_2$, the factor
$T_2^{-1} \sigma_3^{\rm cc}\in {\rm OSp}(2|2)$ can be absorbed in
$T_1$, i.e. the Goldstonemode manifold is isomorphic to a {\it single}
copy of ${\rm Osp}(2|2)$ and can be parameterized by the
representative $T$ freely running through this group. Finally, 
disregarding massive fluctuations (i.e. field-configurations
non-commutative with $\hat H$) and using that the weight factor $
0=\sum_{ij} {\rm \,str\,}(\bar Q_{0,ij}^2+\bar Q_{3,ij}^2)$ is
invariant under transformation by $T$, we identify
\begin{eqnarray*}
Z_{\rm g}=\int {\cal
    D}T  \exp -{1\over 2}{\,\rm str\; ln\,}\left(
 \epsilon \sigma_3^{\rm cc} + \left(\matrix{i\kappa T &-\hat H^{\rm p}_{12}\cr
     -\hat H^{\rm p}_{21} & i\kappa T^{-1}}\right)\right),
\end{eqnarray*}
as an effective functional capturing the low energy degrees of
freedom of the problem.

We next have to face up to the analysis of the pointlike disorder
included in $\hat H^{\rm p}$. More specifically, the object we need to
compute is $Z_{\rm gp} \equiv \langle Z_{\rm g}\rangle_{\rm p}$,
expanded in slow fluctuations of the soft field $T$. Trading the
average over a fixed number of impurities for a grand canonical
average, it is possible to formulate an exact representation of the
averaged functional\cite{mudry00:_unitar}.  However, transcendental in
the fields $T$, the action of the grand canonical functional does not
appear to be a particularly inviting starting point for subsequent
low-energy expansions.  

Below we will adopt a different and less rigorous averaging scheme.
The basic idea will be to employ elements of the standard
approximations implied in the analysis of dilute systems of
scatterers, notably the self-consistent $T$-matrix approximation, in
the expansion of the effective action.
%
%
To be more specific, what we want to do is expand the logarithm
appearing in the action of $Z_{\rm g}$ in (a) the 'mass' brought about
by finite energy arguments $\epsilon$ and (b) the finite action due to
spatial fluctuations of the fields.  The expansion of the
logarithm in these perturbations, for a fixed realization of the
impurity potential, leads to expressions like
$$
Z = \int {\cal D}T \;\left \langle \int d^2 r_1{\cal O}_1(T({\bf r}_1)
  \int d^2 r_2 {\cal O}_2(T({\bf r}_2) \dots
\right \rangle_{\rm p},
$$
where ${\cal O}_{i}(T)$ symbolically stands for a {\it local}
operator depending on the fields $T$ and a continuum approximation is
understood. (Here, the attribute 'local' means locality on the scale
of the elastic mean free path set by the scattering potential.) To
make further progress with this expression, we rely on one central
assumption: For a typical configuration of coordinates $({\bf
  r}_1,{\bf r}_2, \dots)$, the spacings $|{\bf r}_i-{\bf r}_j|$ will
be much larger than the correlation range $\xi$ of the impurity
potential.  For such configurations, the average can be assumed
uncorrelated, i.e.
\begin{eqnarray*}
&&\langle {\cal O}_1(T({\bf r}_1)
 {\cal O}_2(T({\bf r}_2)\dots \rangle_{\rm p}\Big|_{|{\bf r}_i -
{\bf r}_j|\gg \xi} = \\
&&\hspace{1.0cm} 
 = \langle {\cal O}_1(T({\bf r}_1)\rangle_{\rm p}
\langle  {\cal O}_2(T({\bf r}_2)\rangle_{\rm p}.
\end{eqnarray*}
In a small subset of the integration volume, one or several
integration variables come close to each other and averaging over
impurity coordinates leads to residual correlations. This leads to the
generation of composite operators, i.e. local operators
of higher order in gradients and/or energies $\epsilon$. Relying on
the fact 
that
operators of
this type will carry negative scaling dimension, we neglect statistical
correlations and  approximate the average
by 
$$
Z \approx \int {\cal D}T \; \int d^2 r_1\, \langle{\cal O}_1(T({\bf
  r}_1)\rangle_{\rm p} \int d^2 r_2\, \langle {\cal O}_2(T({\bf
  r}_2)\rangle_{\rm p} \dots,
$$
which amounts to replacing the average of the functional by an 
average of the action:
\begin{eqnarray}
\label{eq:22}
&&Z_{\rm gp} \approx \int {\cal D}T\times\\
&&\hspace{.2cm}\times  \exp -{1\over 2}\left \langle {\,\rm str\; ln\,}\left(
 \epsilon \sigma_3^{\rm cc} + \left(\matrix{i\kappa T &-\hat H^{\rm p}_{12}\cr
     -\hat H^{\rm p}_{21} & i\kappa
     T^{-1}}\right)\right)\right\rangle_{\rm p}.\nonumber 
\end{eqnarray}
We will
re-assess the validity of this approximation below.

The rest of the derivation is straightforward. We begin by computing
the action $S_\epsilon$ corresponding to finite values of the energy
parameter $\epsilon$. Expanding the action to first order in
$\epsilon$ and using that for sufficiently slow fluctuations the
fields $T$ and the Hamiltonian $\hat H^{\rm p}$ are approximately
commutative, one obtains
$$
S_\epsilon = -{\epsilon\over 2} {\,\rm str\,} \left(
T \bar G_{11} + T^{-1} \bar G_{22}\right),
$$
where $\bar G = \langle \hat G \rangle_{\rm p}$,
\begin{equation}
  \label{eq:16}
  \hat G= \left(\matrix{i\kappa  &-\hat H^{\rm p}_{12}\cr
     -\hat H^{\rm p}_{21} & i\kappa}
     \right)^{-1}
\end{equation}
is   the disordered Gorkov Green function and
indices are in {\rm ph}-space.
 
\begin{figure}[hbt]
\centerline{\epsfxsize=3.5in\epsfbox{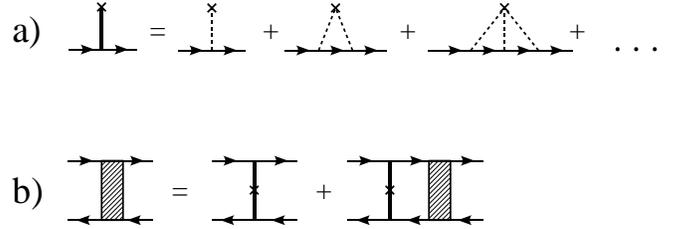}}\vspace{0.5cm}
\caption{a) Diagrammatic representation of the SCTA series for the
  single particle Green function. b) The SCTA diffusion mode.}
\label{fig:SCTA}
\end{figure}\vspace{0.5cm}

Borrowing from previous analyses\cite{lee93}, we compute the
averaged superconductor Green function $\bar G$ by SCTA summation
techniques\cite{fn5}.  The
basic presumption underlying the SCTA (For a more comprehensive
discussion of the approach and its application to the $d$-wave
superconductor see, e.g., Ref. \cite{durst00}.) is that for decreasing
impurity concentration, $n_i$, correlations between scattering
processes off different impurities do not affect the structure of the
averaged Green function.  Neglecting any amount of residual inter-impurity
correlation, the SCTA approximates the self energy, $\hat \Sigma$, of
the Green function by the set of diagrams depicted in Fig.
\ref{fig:SCTA}. Summation over these contributions leads to
$$
\hat \Sigma = n_i \hat T,
$$
where the $T$-matrix (not to be confused with the slow fields $T$ of
the effective action approach) is defined through the Dyson equation 
\begin{eqnarray}
\label{eq:17}
&&  \hat T = v \sigma_2 +  v g \sigma_2 \hat T \Rightarrow\nonumber\\
&&\hspace{0.5cm}\Rightarrow \hat T = (v^{-1}\sigma_2 -  g)^{-1}.
\end{eqnarray}
Here, $ g\equiv \sum_{\bf k} \bar G({\bf k})$, 
\begin{equation}
  \label{eq:24}
\bar G({\bf k}) = \left[i\kappa - \xi({\bf k}) \sigma_2 -
  \Delta({\bf k}) \sigma_1 - \hat \Sigma\right]^{-1},  
\end{equation}
and we anticipate that for a short range scattering potential
$v(k)\equiv v={\rm const.}$, the self energy is momentum independent.
Expanding in Pauli matrices, $ \hat \Sigma = \Sigma_0 + \Sigma_2
\sigma_2$, and defining renormalized parameters $\tilde \kappa \equiv
\kappa + i \Sigma_0\approx i \Sigma_0$ and $\tilde \xi \equiv \xi +
\Sigma_2$, the integrated Green function $g\equiv g_0 + g_2 \sigma_2$
assumes the form
\begin{eqnarray}
\label{eq:19}
g_0= \Sigma_0 \Xi,\qquad
g_2=(\mu-\Sigma_2)\Xi,
\end{eqnarray}
where 
\begin{eqnarray*}
&&  \Xi \equiv  \sum_{\bf k} {1\over  - \Sigma_0^2 +
  \tilde \xi({\bf k})^2 + \Delta({\bf k})^2}.
\end{eqnarray*}
Linearizing the dispersion $H_0({\bf k})$ around the four nodal
points of the Fermi surface of the $d$-wave
superconductor one finds
\begin{eqnarray}
  \label{eq:18}
&&  \Xi \simeq
 {8\over \pi t\Delta} \ln\left(\Lambda\over |\Sigma_0|\right),
\end{eqnarray}
where $\Lambda\gg \Sigma_0$ is a cutoff measuring the extent of the
'linearization volume' in $k$-space. With (\ref{eq:17}), and
$$
\hat T = T_0  + T_2 \sigma_2,
$$
we finally arrive at the self consistency equation
\begin{eqnarray}
  \label{eq:21}
  T_0 &=& {g_0\over -g_0^2 + (v^{-1} - g_2)^2},\nonumber\\
  T_2 &=& {v^{-1} - g_2 \over -g_0^2 + (v^{-1} - g_2)^2}, 
\end{eqnarray}
where $g_{0,2}$ depend on $T_i$ through (\ref{eq:19}).
There are two limiting cases, in which the solution of these equations
is straightforward: In the {\it Born limit}, $v^{-1}-g_2 \gg |g_0|$,
$$
T_0 \stackrel{{\rm Born}}{\longrightarrow} v^2 g_0,\qquad T_2 \stackrel{{\rm Born}}{\longrightarrow} v,
$$
i.e. $T_0$ is given by the familiar first order Born scattering
rate while $T_2$ is the averaged potential background. (Here we have
neglected the chemical potential and the real part of the self energy
for simplicity.)  In the opposite, {\it unitary limit}, $v^{-1}-g_2
\to 0$,
$$
T_0 \stackrel{{\rm unitary}}{\longrightarrow} -g_0^{-1},\qquad T_2
\stackrel{{\rm unitary}}{\longrightarrow} 0.
$$
With these results in place, we turn back to the analysis
of $S_\epsilon$. Expressing the trace over the average Green function
through (\ref{eq:19}) and (\ref{eq:18}) one obtains 
\begin{eqnarray}
\label{eq:27}
&&S_\epsilon =\\
&&\hspace{.3cm}={\epsilon\over 2}\int d^2r \int d^2k  {\,\rm str} \left(
T({\bf r}) \bar G_{11}({\bf k}) + T^{-1}({\bf r}) \bar G_{22}({\bf
  k})\right)=\nonumber\\
&&\hspace{1.5cm} =-i {\pi \epsilon \nu_0\over 4}
\int d^2r   {\,\rm str} \left(
T   + T^{-1}\right)\nonumber,
\end{eqnarray}
where
\begin{equation}
  \label{eq:23}
  \nu_0 = -{1\over \pi} {\,\rm
    Im\,str\,}(g)=
{8 |\Sigma_0|\over \pi^2 t \Delta}{\rm \,ln\,}\left({\Lambda \over
    |\Sigma_0|}\right) 
\end{equation}
is the SCTA density of states at zero energy.  

We next consider the action $S_{\rm fl}$ due  to spatial
fluctuations of the fields $T$. Turning back to (\ref{eq:22}) and
temporarily setting $\epsilon=0$ we use the 
cyclic invariance of the trace, to re-write the action as
$$
S_{\epsilon=0}[T]= {1\over 2} \left \langle {\,\rm str\; ln\,}\left(
 \matrix{i\kappa  &-\hat H^{\rm p}_{12}- T^{-1}[\hat H^0_{12},T]\cr
     -\hat H^{\rm p}_{21} & i\kappa}
\right)\right\rangle_{\rm p}.
$$
(Notice that the impurity potential, diagonal in real space, commutes
with the fields.)  Due to $[H_{12}^0,T] = -i{\bf v}_{12}({\bf k})\cdot
\nabla T + {\cal O}(\partial^2 T)$, where ${\bf v}_{12}({\bf k})
\equiv \nabla_{\bf k}H_{12}$, the commutator between $H_0$ and $T$
contains derivatives of the slow fields and, therefore, counts as a
small object. Expansion of the action to second order (The first order
contribution vanishes by symmetry.) in the commutator leads to
$$
S_{\rm fl} = {1\over 4}\left \langle {\rm str}\left[\hat G_{21}  
(T^{-1}{\,\bf v}_{12} \cdot \nabla T) \,
\hat G_{21}(T^{-1} {\,\bf v}_{12} \cdot
\nabla T)\right]\right\rangle_{\rm p}.
$$
In the limit of negligibly small real space correlation of the
potential (assumed throughout), $v({\bf k})\equiv v_0={\rm \, const.}$, vertex corrections to
$S_{\rm fl}$ vanish, i.e.
$$
S_{\rm fl} = {1\over 4} {\rm str}\left[\bar G_{21}  
(T^{-1}{\,\bf v}_{12} \cdot \nabla T) \,
\bar G_{21}(T^{-1} {\,\bf v}_{12} \cdot
\nabla T)\right].
$$
At this stage, the further evaluation of $S_{\rm fl}$ seems
straightforward: One should substitute (\ref{eq:24}), adopt the linear
('nodal') approximation previously used in (\ref{eq:18}), and do the
momentum integral. However, there is a subtle problem with this
strategy. Straightforward substitution of the linearized dispersion
leads to expression of the type $0\cdot \infty$, where $'0'$ is due to
the $k$-space symmetry of the integral and $'\infty'$ due to
ultraviolett divergencies. 

There are different ways to get around this problem. Abandoning the
nodal approximation one may attempt to do the full momentum integral
over the Brioullin zone. This integral is UV benign (We are working on
a lattice!) but cumbersome. Alternatively, one may stick to the nodal
approximation and implement one of several UV regularization schemes
available for fermions with linear dispersion. Choosing the second
route, and following a regularization procedure introduced in Ref.
\cite{altland00.1}, we obtain
\begin{equation}
  \label{eq:25}
  S_{\rm fl}[T] = - {D \nu_0 \pi \over 8} \int d^2 r {\,\rm
    str\,}(\partial T \partial T^{-1}), 
\end{equation}
where the -- disorder independent -- coupling constant is related to
the parameters $t$ and $\Delta$ through (\ref{eq:29}).  Combination of
Eqs.  (\ref{eq:27}) and (\ref{eq:25}) finally yields the effective low
energy action (\ref{eq:3}) discussed in the introduction.

At this stage, it is instructive to make explicit contact with the
perturbative analysis of YAGHK. To this end, we expand the fields as
$T=\exp(W) = \openone + W + W^2/2 + \dots$, where $W\in {\rm
  osp}(2|2)$ are group generators of ${\rm OSp}(2|2)$.  Expansion of 
the action to second order in $W$ then obtains
\begin{eqnarray}
\label{eq:41}
  S[W] = {1\over 2\pi} 
\int d^2r {\,\rm str}\left(W \left(- D 
\partial^2 - 2 i  \epsilon  \right)W \right) + \dots,
\end{eqnarray}
where the kernel $\Pi^{-1}\equiv -D\partial^2 + 2 i \epsilon$ is the
inverse of the SCTA particle-hole diffusion mode depicted in Fig.
\ref{fig:SCTA} b) and derived diagrammatically in YAGHK.  (That the
quadratic action correctly describes diffusion on the background of
the isolated impurity centers is evidence for the validity of the
statistical assumptions underlying the derivation of the action.)  To
compute the first order quantum interference corrections to the DoS,
one expands the functional expectation value (\ref{eq:4}) to second
order in $W$ and integrates.  This produces the logarithmically
divergent results of YAGHK. For a list of more
sophisticated schemes of evaluating the field theory, we refer back to
the discussion of \ref{sec:qual-disc-results}.

\section{The Half-Filled Band}
\label{sec:half-filled-band}

Building on the analysis of the previous section, we next consider the
consider the unitary limit (by our abuse of language, the combined
limit half filling/infinitely strong scattering.) To understand what
will change, let us go back to the level of the prototypical field
theory (\ref{eq:20}) and consider the bi-linear form
$$
\bar \Psi \hat H_0 \Psi.
$$
As discussed in section \ref{sec:qual-disc-results}, the clean
Hamiltonian $\hat H_0$ possesses the symmetries ${\cal C}, {\cal T}$,
{\it and} the sublattice symmetry ${\cal N}$.
As we are aiming to describe a situation where not only ${\cal C}$ and
${\cal T}$, but also ${\cal N}$ is left intact, let us explore how this
symmetry will change the internal structure of the field theory. 

As discussed in section \ref{sec:qual-disc-results}, the 
symmetry ${\cal N}$ is linked to the existence of two nested
sublattices $A$ and $B$. A superconductor Hamiltonian 
describing hopping between these lattices can be symbolically represented as
\begin{equation}
  \label{eq:34}
  \hat H_0 = \hat h_1 \otimes \sigma_1 + \hat h_2 \otimes \sigma_2,
\end{equation}
where the lattice Hamiltonians $\hat h_i$ obey $[\hat h_i,
\sigma_3^{\rm sl}]_+=0$, $i=1,2$, and $\sigma_3^{\rm sl}$ is defined
in (\ref{eq:31}). Likewise, the fields $\Psi$ and $\bar \Psi$ carry
a two-component structure in sub-lattice space. The important
point now is that for a theory globally compatible with 
${\cal N}$, the  internal symmetry is larger than that discussed in
the previous section. Indeed, it it straightforward to verify that
$\bar \Psi \hat H_0 \Psi$ remains invariant under transformations
\begin{eqnarray}
\label{eq:35}
&&  \Psi \to \hphantom{\Psi}\left[\matrix{
T_1 &&& \cr
& T_2 && \cr
& & T_1^{s-1} & \cr
&&& T_2^{s-1}}\right]\Psi\equiv {\rm T}  \Psi,\\
&&
\bar \Psi \to \bar \Psi \left[\matrix{
T_2^s &&& \cr
& T_1^s && \cr
& & T_2^{-1} & \cr
&&& T_1^{-1}}\right]\equiv \bar \Psi {\rm \bar{T}},
\end{eqnarray}
where the matrix structure in sublattice and particle-hole space is
defined through $(A{\rm p},A{\rm h},B {\rm p}, B {\rm h})$ and
$T_{1,2}$ are two {\it independent} matrices drawn from ${\rm
  GL}(2|2)\supset {\rm OSp}(2|2)$. In other words, the symmetry of the
theory has grown from ${\rm OSp}(2|2)\times {\rm OSp}(2|2)$ to the
larger manifold ${\rm GL}(2|2) \times {\rm GL}(2|2)$.

Having understood that, we proceed to explore (i) the structure of the
fields entering the functional integral, (ii) the form of the low
energy action, (iii) the role of finite chemical potentials $\mu$,
and, most importantly, (iv) why the symmetry ${\cal N}$ survives
generalization to unitary disorder.

Temporarily ignoring the Poisson disorder, we notice that, as before, the
diagonal configuration $i \kappa \sigma_3^{\rm cc}$ solves the saddle
point equation. Transformation by a smoothly fluctuating field of the
structure (\ref{eq:35}), then leads to
\begin{equation}
  \label{eq:26}
  i\kappa {\rm T} \sigma_3^{\rm cc} {\rm \bar{T}}\equiv 
i \kappa \left[\matrix{
T&&&\cr
&T^s &&\cr
&&T^{s-1}&\cr
&&&T^{-1}}\right],
\end{equation}
which identifies $T\equiv T_1 \sigma_3^{\rm cc} T_2^s\in {\rm GL}(2|2)$ as  the
soft field of the theory and answers (i). 

Turning to (ii), we note that with the results of the previous section
in place, the low energy structure of the action can be fixed by
symmetry considerations, plus a few algebraic manipulations. Rather
than explicitly expanding in slow fluctuations of the field $T$, we
use that the low-energy action $S[T]$ we are looking for must collapse
to (\ref{eq:3}) if the symmetry ${\cal N}$ is broken. This condition
alone implies that the action of the system is given by (\ref{eq:32}).
Indeed, for fields drawn from the subset ${\rm OSp}(2|2)\subset {\rm
  GL}(2|2)$ (i.e. the transformation group of the non-${\cal
  N}$-symmetric model) $S[T]$ reduces to the action derived in the
previous section. Further, $S[T]$ contains all two-derivative
operators compatible with the gross symmetries of the system.  The
only coupling constant that cannot be fixed by simple symmetry
reasoning is $c$. However, for the purposes of the present analysis,
the bare value of this constant is inessential.

We next ask, (iii), how finite values of the chemical potential $\mu$
affect the picture. For finite $\mu$, and still ignoring the presence
of the Poissonian disorder, the intermediate action assumes the form
(cf. the fourth line in Eq. (\ref{eq:20}))
$$
{\rm str\;\ln}\left(i\kappa {\rm T}\sigma_3^{\rm cc}{\rm \bar T} + \mu
  \sigma_1 - \hat H_0\right).
$$
One verifies that the isolated Pauli matrix $\sigma_1$ is
incompatible with the full set of transformations encapsulated in
${\rm T}$; using the language of (\ref{eq:26}), only transformations
with $T^s=T^{-1} \Rightarrow T\in {\rm OSp}(2|2)$ commute with
$\sigma_1$.  To quantitatively describe the symmetry crossover, one
expands the action to second order in the symmetry breaking parameter
$\mu$. (The first order term vanishes after integration over the
eigenvalues of $\hat H_0$.) This obtains the symmetry breaking
operator (\ref{eq:33}), with coupling
\begin{equation}
  \label{eq:40}
  \Gamma \equiv \Gamma_1 \sim {\mu^2 \over t \Delta}.
\end{equation}

We finally, (iv), need to address the question of disorder scattering.
The generic Poisson potential $V^{\rm p}$ (\ref{eq:36}) is clearly
incompatible with the lattice symmetry ${\cal N}$. However, for a
singular potential, strong enough to locally eradicate all
wave-function amplitudes in the low-energy sector of the Hilbert
space, there is a chance for ${\cal N}$ to remain {\it effectively}
intact. An infinitely strong local potential punches holes (sites where
all wavefunctions vanish) into the system. Multiplication of these
sites by a phase factor $\pm 1$ (the action of $\sigma_3^{\rm sl}$) is
a redundant operation, i.e. ${\cal N}$ still holds. 

To quantitatively verify that this exceptional scenario is realized
for unitary scattering at half filling, we consider the Green
function operator
$$
\hat G[T] = \left[i\kappa {\rm T}\sigma_3^{\rm cc}{\rm \bar T} 
  - \hat H^{\rm p}\right]^{-1}
$$
as the central building block entering the expansion of the action.
Notice that $\hat G[T]$ differs from the 'bare' Green function $\hat
G$, Eq. (\ref{eq:16}), in that the soft fields are included.  The
effect of impurity scattering on the local spectral weight is
described by the self energy operator $\hat \Sigma[T]$ of the averaged
Green function 
\begin{equation}
  \label{eq:37}
\bar G[T]=\left[i\kappa {\rm T}\sigma_3^{\rm cc}{\rm  
    \bar T} - \hat H_0 - \hat \Sigma[T]\right]^{-1}.
\end{equation}
To understand the connection between the impurity strength and the
invariance properties of the Green function, let us consider the self
energy $\Sigma_i[T]$ due to a {\it single} local impurity potential
$v_i$ located at site $i$. (Effectively, this amounts to neglecting
the effect of inter-impurity correlations on the wave-function
amplitudes at individual scattering sites.)  Solution of the, now
exact, Dyson series
$$
\Sigma_i[T] = v_i \sigma_2 + v_i\sigma_2 \bar G[T]_{ii} \Sigma_i[T]
$$
leads to 
\begin{equation}
  \label{eq:38}
\Sigma_i[T] = \left[v_i^{-1}\sigma_2- \bar G[T]_{ii}
\right]^{-1}.   
\end{equation}
We next employ the $T$-matrix equations (\ref{eq:37}) and
(\ref{eq:38}) to explore the invariance properties of the theory under
transformations $T$. Due to ${\rm T} \hat H_0 \bar{\rm T}^{-1}= \hat
H_0$, the non-disordered Green function $\hat G_0[T]$ transforms as
$\hat G_0[T] = \bar{\rm T}^{-1} \hat G_0[\openone] {\rm T}^{-1}$. Had
the disordered Green function the same property, the theory would be
invariant under the full set of transformations $T$. From Eq.
(\ref{eq:37}) we see that invariance of the Green function is ensured
if $\Sigma_i[T] = {\rm T} \Sigma[\openone] \bar {\rm T}$.  But if the
Green function {\it was} invariant, we would have,
\begin{eqnarray*}
\lefteqn{\Sigma_i[T]=\left[v_i^{-1}\sigma_2- \bar{\rm T}^{-1}\bar
  G[\openone]_{ii}{\rm T}^{-1}\right]^{-1}}\hspace{1cm}\\
&&  = {\rm T} \underbrace{\left[v_i^{-1} \bar{\rm T} \sigma_2 {\rm T}- \bar
  G[\openone]_{ii}\right]^{-1}}_{\displaystyle\not=\Sigma_i[\openone]}\bar{\rm T}.
\end{eqnarray*}
The last line shows that for generic values of the impurity strength,
the self energy operator does not have the required invariance
properties. Only in the limit of an infinitely strong potential
$v_i\to \infty$, the offending term $v_i^{-1} \bar{\rm T} \sigma_2
{\rm T}$ drops out and the action is invariant under the full set of
transformations $T$.  This is the unitary limit where the system is
described by the soft action (\ref{eq:32}). (Notice that for
transformations drawn from the subgroup ${\rm OSp}(2|2)$ (i.e. the
symmetry group of the non-${\cal N}$-symmetric superconductor)
$v_i^{-1} \bar{\rm T} \sigma_2 {\rm T}=v_i^{-1} \sigma_2$, i.e.
invariance under the reduced set of transformations holds regardless
of the impurity strength.)

For scattering close to the unitary limit, $v \gg 1$, an expansion of
the action to second order in $v^{-1}$ obtains, again, the operator
(\ref{eq:33}), this time with coupling 
\begin{equation}
  \label{eq:42}
  \Gamma \equiv \Gamma_2 \sim  
{n_i\over (\nu_0 v)^2}.
\end{equation}
The operator (\ref{eq:33}), with coupling $\Gamma = \Gamma_1 +
\Gamma_2$, describes the breaking of $A$III symmetry by finite
chemical potentials and/or deviations off unitarity. An expansion of
(\ref{eq:33}) to second order in the field generators $W$ (cf. Eq.
(\ref{eq:41})) shows that for finite $(\mu,v^{-1})$ the soft mode (the
'diffuson') induced by the sublattice symmetry contains two mass
terms, derived and discussed before by YAGHK. Whether or not traces of
the $A$III-type soft modes remain visible in the long range behaviour
of transport and spectral observables depends on the relative value
$\Gamma/(\nu_0 \epsilon)$. If this parameter is small, the mass of the
superconductor soft modes is mainly set by deviations off zero energy
and the (relatively small) additional breaking of the sublattice
symmetry is inessential. In the opposite case, the sublattice symmetry
is strongly broken and the system effectively in class $C$I.

\section{Summary}
\label{sec:summary-discussion}
This concludes our field-theoretical analysis of $d$-wave
superconductors with pointlike scatterers.  We have shown that in all
but one point of a phase plane defined through impurity strength and
filling factor, respectively, the system behaves similar to $d$-wave
superconductors with generic, non-pointlike disorder. The DoS vanishes
linearly upon approaching zero energy and eigenstates are localized
(i.e. the system is a thermal insulator.)

There exists one point in the phase diagram -- infinitely strong
disorder and half filling -- where phenomenology of strikingly
different type is realized: The DoS diverges as $\nu(\epsilon) \sim
|\epsilon|^{-1} f(\ln|\epsilon|)$, where $f(\ln|\epsilon|)$ stands for
logarithmic corrections to power law scaling\cite{pepin98}, and
zero-energy eigenstates are localized.   Apart from deviations in the
function $f$ -- whose origin remains obscure -- the spectral profile
found in the present analysis agrees with that of Ref.
\cite{pepin98}. The identification of a metallic phase is a new
result.

We do not agree with the statement\cite{pepin98} that the
peculiar behaviour realized at the critical point should be by and
large insensitive to detuning of the parameters chemical
potential/impurity strength. 
The question whether traces of critical behaviour -- e.g. regions
where the DoS increases upon lowering the energy -- might be
observable under 'realistic conditions' is not for the present
approach to decide. However, given that the very formation of a
superconducting phase requires a finite concentration of dopands (i.e.
deviations off half-filling), and that residual quasiparticle
interactions spoil the sublattice symmetry, it seems likely that the
'real' superconductor will display behaviour characteristic for the
generic symmetry class.\\

{\it Note added:} Shortly before completion of this manuscript
Ref.\cite{fabrizio} appeared. The authors of that reference discuss
structure and renormalization of superconductor
$\sigma$-models with sublattice symmetry 
(including the case of broken time reversal invariance.) Where the
same symmetry classes are considered, the results of Ref. \cite{fabrizio}
and the present paper agree. 

It is a pleasure to acknowledge discussions with
  C. Mudry, B. D. Simons, and M. Zirnbauer.
\appendix

\section{Block Representation of the Superconductor with Nesting
  Symmetry}
\label{sec:block-repr-superc}

We show that the Hamiltonian $\hat H$ of the superconductor with nesting
symmetry can be transformed to a block off-diagonal representation.
Due to the symmetry (\ref{eq:31}), $\hat H$ can be represented through
(\ref{eq:44}), where the matrix structure is in sublattice
space. The superconductor symmetries (\ref{eq:1}) further imply that 
the block-matrices $\hat Z$ obey
$$
Z=Z^\ast, \qquad \sigma_y Z \sigma_y = -Z,
$$
i.e. $\hat H$ has the form
\begin{eqnarray*}
  \hat H = \left[\matrix{
& & h_2 & h_1\cr
& & h_1 & -h_2\cr
h_2^T & h_1^T &&\cr
h_1^T & -h_2^T &&}\right],
\end{eqnarray*}
where the sub-structure of the blocks is in ${\rm ph}$-space and the
operators $h_{1,2}$ are real.  A unitary transformation in ${\rm
  ph}$-space brings $\hat H$ into the form
\begin{eqnarray*}
  \hat H = \left[\matrix{
& & & \tilde Z \cr
& & \tilde Z^\ast & \cr
 & \tilde Z^T &&\cr
\tilde Z^\dagger &  &&}\right],
\end{eqnarray*}
where we have defined the complex operator $\tilde Z = h_1 - i
h_2$. Finally, changing the order of blocks, $\hat H$ can be
rewritten as
\begin{eqnarray*}
  \hat H = \left[\matrix{
&\tilde Z & &  \cr
\tilde Z^\dagger & &  & \cr
 &  &&\tilde Z^\ast\cr
 &  &\tilde Z^T&}\right].
\end{eqnarray*}
(Apart from an inessential transposition in the lower left hand
block,) these are two identical copies of a class
$A$III-Hamiltonian. The replicated appearance of  {\it two}
$A$III-Hamiltonians implies that our field theory takes values in the
doubled field manifold ${\rm GL}(2|2)$ (instead of ${\rm GL}(1|1)$ for
just one Hamiltonian.)

\end{multicols}

\begin{references}
\bibitem{fn1} Superimposed on the point-like system of scatterers,
  there may be softer disorder potentials caused by the presence of
  defects in the inter-planar regions.
\bibitem{fn3}
Notice that for $T\in {\rm OSp}(2|2)$, the spurious
  third term in (\ref{eq:32}) vanishes (because ${\, \rm
    str\,}(T\partial T^{-1}) = \partial {\,\rm str\,}(\ln T)= \partial
  {\,\rm ln \; sdet\,}(T)$, and the fact that ${\rm OSp}(2|2)$ has
  superdeterminant 1.) The rest of the action coincides with
  (\ref{eq:3}).
\bibitem{fn4}
Of course, the
  structure of the model must be insensitive to the way by which its
  low energy degrees of freedom are introduced.  E.g., avoiding the
  introduction of Gaussian disorder, one might equally well construct
  the theory by averaging over a window of chemical potentials.
  Technically, however, it is more convenient to implement the
  disorder-route.
\end{references}

\end{document}